# Improving Survival Models in Healthcare
# by Balancing Imbalanced Cohorts:
# A Novel Approach

Catherine Ning[1], Dimitris Bertsimas[1], Johan Gagnière[2], Stefan Buettner[3], Per Eystein Lønning[4], Hideo Baba[5], Itaru Endo[6], Georgios Stasinos[7], Richard Burkhart[8], Federico N. Auecio[9], Felix Balzer[10], Cornelis Verhoef[3], Martin E. Kreis[11], Georgios Antonios Margonis[1] [12]

**Affiliations**
[1] Operations Research Center, Massachusetts Institute of Technology, Cambridge, MA, USA
[2] Department of Digestive and Hepatobiliary Surgery–Liver Transplantation U1071 Inserm/Clermont-Auvergne University Hospital of Clermont-Ferrand, Clermont-Ferrand, France
[3] Department of Surgery, Erasmus MC University Medical Centre, Rotterdam, The Netherlands
[4] Department of Clinical Science, University of Bergen, Department of Oncology, Haukeland University Hospital, Bergen, Norway
[5] Department of Gastroenterological Surgery, Graduate School of Medical Sciences, Kumamoto University, Kumamoto, Japan
[6] Department of Gastroenterological Surgery, Yokohama City University Graduate School of Medicine, Yokohama, Japan
[7] Technical Chamber of Greece, Athens, Greece
[8] Department of Surgery, Johns Hopkins University School of Medicine, Baltimore, MD, USA
[9] Department of General Surgery, Digestive Disease Institute, Cleveland Clinic, Cleveland, OH, USA
[10] Charité – Universitätsmedizin Berlin, Berlin, Germany
[11] Department of General and Visceral Surgery, Charité Campus Benjamin Franklin, Berlin, Germany
[12] Department of Surgery, Memorial Sloan Kettering Cancer Center, New York, NY, USA

**Corresponding Author**
Catherine Ning,

Ph.D. candidate in Operations Research Center, MIT
Email: cat2510@mit.edu

**100-word Context Summary**

This work addresses a critical yet underexplored challenge in biomedical prognostication: how imbalanced representation across prognostic subgroups in observational cohorts can undermine survival model performance in underrepresented tail groups. Our novel methodology rebalances the risk distribution of training datapoints to shift the model's focus toward the low- and high-risk tail groups, thereby improving prognosis in these clinically important patients. We developed our method using a real-world cohort of patients with colorectal liver metastases and validated it in two external, independent cohorts. On external validation, we found considerable gains in Harrell's C index within the underrepresented high- and low-risk strata.

**Abstract**

**PURPOSE:** We explore whether survival model performance in underrepresented high- and low-risk subgroups—regions of the prognostic spectrum where clinical decisions are most consequential—can be improved through targeted restructuring of the training dataset. Rather than modifying model architecture, we propose a novel risk-stratified sampling method that addresses imbalances in prognostic subgroup density to support more reliable learning in underrepresented tail strata.

**MATERIALS AND METHODS:** We introduce a novel methodology that partitions patients by baseline prognostic risk and applies matching within each stratum to equalize representation across the risk distribution. We implement this framework on a cohort of 1,799 patients with resected colorectal liver metastases (CRLM), including 1,197 who received adjuvant chemotherapy and 602 who did not. All models used in this study are Cox proportional hazards models trained on the same set of selected variables. Model performance is assessed via Harrell's C index and Integrated Calibration Index (ICI), with internal validation using Efron's bias-corrected bootstrapping. External validation is conducted on two independent CRLM datasets.

**RESULTS:** Cox models trained on risk-balanced cohorts showed consistent improvements in internal validation compared to models trained on the full dataset. The proposed approach preserved overall model calibration while noticeably improving stratified C-index values in underrepresented high- and low-risk strata of the external cohorts.

**CONCLUSION:** Our findings suggest that survival model performance in observational oncology cohorts can be meaningfully improved through targeted rebalancing of the training data across prognostic risk strata. This approach offers a practical and model-agnostic complement to existing methods, especially in applications where predictive reliability across the full risk continuum is critical to downstream clinical decisions.

# Introduction

Prognostication plays a central role in biomedical research, with numerous prognostic models already being employed in clinical practice to inform patient outcomes and guide decision-making [1]. Conventionally, prognostic models are trained on full observational datasets under the assumption that the naturally occurring distribution of patient characteristics and risk levels is optimal for model learning.

We challenge this assumption by asking: when certain *prognostic risk groups* are *disproportionately represented* in the training data, how does this impact model learning and generalization? We hypothesize that such "risk distribution imbalance" can bias the model to focus on risk strata with high sample density. This is particularly problematic when the underrepresented strata are at the extremes of risk—precisely where treatment decisions are most critical and the stakes highest. For example, in the context of colorectal liver metastases (CRLM), patients with very high-risk disease (e.g., large or multifocal tumors) may be considered for intensified perioperative adjuvant chemotherapy, whereas patients with very low-risk disease (e.g., small, biologically favorable tumors) may undergo surgery alone and avoid the toxicity of unnecessary systemic therapy.

We introduce a novel methodology to construct balanced training datasets to amplify signals from the originally underrepresented risk strata during model learning. First, patients are stratified by their estimated five-year mortality risk, then, within each risk subgroup, we solve an optimal matching problem to undersample dominant strata. This yields a new training dataset that is uniformly distributed across low-, medium-, and high-risk strata while preserving balance between treatment groups.

We apply this methodology to a multi-institutional dataset of patients with CRLM treated at tertiary academic centers, where most received adjuvant chemotherapy. We compare the resulting models to a baseline model trained on the original cohort with Efron's bias-corrected bootstrap for internal validation [3] and with external validation in two independent CRLM cohorts. Across sensitivity analyses, the Cox models trained using our methodology consistently outperformed the standard training approach without rebalancing and most importantly, achieve stronger C-indices in the "tails" of both external datasets (i.e., very low- and high-risk patients). This suggests that balancing can systematically enhance model performance precisely where clinical decision-making is most critical.

# Methods

## Model Specifications

The Cox proportional hazards (CPH) model is the most widely utilized learning algorithm for survival analysis, including studies for CRLM [2,4], thus all the models are trained using Cox regression on the same prognostic factors. The prediction target is the five-year overall survival, standard in oncological studies. Furthermore, to remain consistent with the study cohort (see Results), we will refer to the two treatment arms as "surgery + chemotherapy" and "surgery alone". We define:

- **Model 1**: CPH model trained on the entire cohort of the original dataset, which is the standard approach in prognostic models for CRLM patients [2].

- **Model 2A:** CPH model trained exclusively on the "surgery alone" subcohort and **Model 2B** exclusively on the "surgery + chemotherapy" subcohort. Comparing these to Models 3A and 3B allows us to isolate the effect of risk stratified balancing.

- **Model 3A**: CPH model trained on the balanced "surgery alone" subcohort and **Model 3B** on the balanced "surgery + chemotherapy" subcohort after applying our methodology.

## Prognostic Stratum Matching

Motivated by our hypothesis that imbalanced sample densities in prognostic subgroups negatively affect model performance, we propose a matching-based approach to undersample dominant risk subgroups in the training dataset. The goal is not causal effect estimation, but rather to equalize representation across the spectrum of baseline prognostic risk. By balancing strata, we enable Cox Models 3A and 3B to learn treatment-specific prognostic signals more effectively across the entire risk spectrum, including the clinically important but sparsely populated high- and low-risk tails.

Our prognostic stratum matching algorithm comprises two stages:

1. stratify the training data by estimated five-year prognostic risk;
2. equalize risk stratum sizes using optimal matching between treatment arms.

In the first step, we group patients into prognostic strata according to their baseline prognosis — the predicted five-year mortality risk if patients had not received adjuvant chemotherapy. To estimate the baseline prognostic risk, we fit a CPH model exclusively on the "surgery alone" patients, then perform inference on both "surgery alone" and "surgery + chemotherapy" patients. By discretizing the estimated probabilities into risk strata, we group patients by similar baseline prognosis.

In the second step, within each stratum, we solve an integer optimization problem (Problem1), a one-to-one minimum distance pair matching without replacement between "surgery alone" and "surgery + chemotherapy" patients in the training data. The matching criteria is based on minimizing the pairwise Euclidean distance of the prognostic factors (same as those to fit the survival models). This matching procedure reduces over-representation of dense strata while preserving diversity across the prognostic spectrum. Importantly, we impose no caliper constraints, since stratification already ensures within-stratum comparability. Algorithm 1 in Appendix p1 details the full algorithmic flow.

Formally, within each stratum, let us denote the smaller treatment group as Group $A$, the larger as Group $B$. Problem1 involves binary decision variables $z_{ij}$, with $z_{ij} = 1$ indicating patient $i$ from $A$ is matched to patient $j$ from $B$ and $z_{ij} = 0$ otherwise. The objective is to minimize prognostic differences, quantified by the Euclidean distance of $K$ features between matched pairs. Moreover, we specify a threshold parameter $\alpha$ of minimum matched pairs per stratum. When Group $A$ in a given stratum has at least $\alpha$ datapoints, we undersample the stratum until both Groups $A$ and $B$ reach size $\alpha$ (Constraint Line 3), while Constraints 1 and 2 ensure that each patient is matched at most once. Otherwise, we keep all samples from Group $A$ and match each of them with exactly one closest sample from

Group $B$. This as well as a relaxed problem formulation can be found in Appendix p.2. $\alpha$ was tuned using internal validation (see Supplemental Figure S2).

$$\min_z \sum_{i \in A} \sum_{j \in B} d_{ij} z_{ij}$$

$$\text{s.t.} \quad \sum_{j \in B} z_{ij} \leq 1 \quad \forall i \in A$$

$$\sum_{i \in A} z_{ij} \leq 1 \quad \forall j \in B$$

$$\sum_{i \in A} \sum_{j \in B} z_{ij} \geq \alpha$$

$$d_{ij} = \sqrt{\sum_{k=1}^{K} (\hat{x}_{ik} - \hat{x}_{jk})^2}$$

$$\hat{x}_{ik} = \frac{x_{ik} - \bar{x}_k}{\sigma_k}$$

$$z_{ij} \in \{0,1\}$$

where $\bar{x}_k$ and $\sigma_k$ are the mean and the standard deviation of an unnormalized feature vector $x_k$ across all patients in the dataset.

## Internal and External Validation

To evaluate both discrimination and calibration accuracy of our models, we evaluate three standard metrics in survival analysis: Harrell's C [5], time-dependent AUC [6], and ICI [7].

For internal validation, we use Efron's bias-corrected bootstrapping [3,8,9] technique to provide unbiased estimates of the future performance through repeated model fitting on the training data (before or after balancing) with correction for optimism. In comparison, a conventional train-test-splitting of our study cohort would result in even fewer train and test datapoints in the tail strata, making it impossible to assess improvements in those underrepresented groups (further details and implementation in Appendix pp3-4).

For external validation, Model 1 and Models 3A/B trained on our development cohort are applied unchanged to the external cohorts to generate predictions and report (i) overall metrics and (ii) per-stratum performance to illustrate where along the risk spectrum of the external cohort gains occur. This is because when scores are averaged across the entire test set, they tend to hide performance gains in the underrepresented groups. In contrast, a per-stratum evaluation of our models may reveal targeted improvements in the tail groups where prognostic accuracy matters most. To reveal the risk spectrum specific to the external cohort, we fit a Cox model on the external cohort and use the estimated five-year mortality risks to stratify the external patients. This Cox model is used only for binning and is independent of the evaluated models, thus this procedure does not introduce information leakage nor inflate performance; rather, it provides a cohort-appropriate lens for subgroup reporting. For deployment, no stratification nor any preprocessing except for variable imputation is required to obtain predictions for a new patient: the fixed treatment-specific model is applied directly.

Details on metrics, imputation, and software specifications are in Appendix p.5.

# Results

## Study Cohort

In this observational cohort study, we consider 2,375 adult patients who underwent surgery for CRLM. Patients are included if they have complete records for adjuvant chemotherapy status and overall survival as these variables are deemed inappropriate for imputation if missing. This results in a final cohort of 1,799 eligible patients, as shown in Supplemental Figure S1. After thorough literature review, we select ten prognostic baseline variables [10] (Supplemental Table S1 and Table 1 before and after imputation). The median OS is 66.5 months, with 1-year, 3-year, and 5-year survival rates of 92.33%, 70.20%, and 52.59%, respectively. The median follow-up time is 61.4 (IQR: 35.6–96) months.

## Prognostic Stratum Matching

Our stratification step divides patients into 8 prognostic strata, striking a balance between sufficient resolution of the risk spectrum and intuitive presentation in probability deciles. Figure 1 contrasts the imbalanced prognostic distribution (five-year mortality risk) of the original dataset with that of our balanced dataset. The original cohort is heavily skewed toward the medium-risk strata for both treatment arms, while low- and high-risk patients are underrepresented (tails). 1-1 prognostic stratum matching restructures the dataset into uniform sample sizes across strata and treatment groups[1]. In Appendix, Figure S3 shows the histogram after applying relaxed matching instead of 1-1 matching, Figure S4 the Kaplan-Meier plots before and after balancing, and Table S2 the baseline characteristics of the balanced cohort.

## Internal Validation

Table 2 and Table 3 compare the in-sample respectively bias-corrected bootstrapping performance of the Cox models in this study, highlighting the following observations:

---

[1] solving our integer optimization problem for eight strata requires only 0.8 seconds on average; hence, for datasets of this size, our undersampling method does not incur computational overhead

- Models 3A and 3B, trained on the balanced cohorts after prognostic stratum matching, consistently achieve the highest discrimination metrics with up to 15% higher bias-corrected C indices compared to Model 1 trained on the original imbalanced data.

- Comparing the two matching problem formulations, relaxed matching for Model 3B seems to yield best C-index, whereas 1-1 matching for Model 3A shows superior performance.

- While prognostic stratum matching improves discrimination, its effect on calibration can be more variable.

## External Validation

We hypothesized in our introduction that models trained on balanced cohorts may generalize better, especially in sparse prognostic subgroups, than those trained on imbalanced cohorts. To test this hypothesis, we evaluate the out-of-sample performance of Models 3A and 3B against Model 1 in two independent cohorts.

The first external dataset ("French") includes 660 CRLM patients treated with surgery and chemotherapy at the Departments of Surgery of CHU Clermont-Ferrand and Université Clermont Auvergne, while the second dataset ("Dutch") comprises 1,058 CRLM patients who underwent surgery alone at the Department of Surgery of the University of Rotterdam. Baseline characteristics are in Supplemental Table S3.

Figure 2 indicates that the French cohort is left-skewed, concentrating predominantly in lower-risk regions, with barely any samples above 60% risk of death. The Dutch cohort mirrors the light-tailed distribution of our development cohort in Figure 1. Since the French dataset only contains "surgery + chemotherapy" patients while the Dutch dataset only "surgery alone" patients, we validate Model 3B on the French, and Model 3A on the Dutch cohort .

Due to the small effective number of samples in the tail strata of the external cohorts, there is inevitably larger evaluation uncertainty present in all the models when assessed within these strata. Therefore, to demonstrate robustness of our methodology, we performed sensitivity analyses by varying (i) the number of strata (from 7 to 10) and (ii) testing different subsets of prognostic factors. Across these analyses, the results of which are in Appendix p.12-19, Models 3A/3B outperformed the current gold standard (Model 1). Figure 3 reveals that Model 3B achieves higher C-index scores than with up to 19% improvement within the tails of the French cohort. Similarly, Model 3A outperforms Model 1 within the lowest (0–0.3) and highest (0.8–1.0) risk strata of the Dutch cohort, where the improvement is 32% respectively 14%. Global performance metrics are reported in Supplemental Table S4.

These findings suggest a critical practical implication: training on balanced data can achieve meaningful prognostic improvements precisely in those clinically important low- and high-risk groups.

## Discussion

The problem of imbalance has been extensively studied in machine learning classification, where unequal class representation is known to degrade performance in minority classes [11]. Here, instead of imbalance in discrete outcome classes, we address the critical yet underexplored challenge of imbalance across prognostic risk strata in survival analysis. Although imbalances in baseline risk are common in clinical datasets, most survival models are trained directly on these data without adjustment [2].

At first glance, discarding data during training may appear undesirable. However, we argue that selective removal of data points can be beneficial if it encourages models to learn equally from all prognostic subgroups. To this end, we introduced a novel principled undersampling strategy that balances risk strata by matching and then trains two separate Cox models — one per treatment arm — each exposed equally to low-, medium-, and high-risk patients. Unlike propensity score [12] or prognostic score [13] matching in causal inference, which aim to reduce confounding or redefine the target population, our use of matching serves a different purpose. We deliberately restructure the training data to improve the model's ability to learn prognostic signals across the entire risk spectrum, particularly in the tails. Our internal validation using bias-corrected bootstrapping provides stable and nearly unbiased estimates of performance within the study cohort, especially compared to conventional data splitting, which would leave too few patients in the underrepresented strata for reliable evaluation. Nevertheless, we view these internal estimates as complementary rather than conclusive. To assess true generalizability, we rely on external validation in two independent CRLM cohorts, an approach broadly considered the gold standard because it mirrors how the models would perform in entirely new populations.

In both external cohorts, the models trained on balanced data outperformed the comparator model at the extremes of the risk spectrum with up to 54% relative increase in C-index (from 0.47 to 0.73 in the Dutch leftmost stratum). Evaluating model performance within individual risk strata is a novel approach that highlights targeted improvements in underrepresented patient groups, which are often obscured by global indices averaged across the entire dataset. While the tail groups may have limited population-level impact due to their small patient count, these accuracy gains are clinically meaningful, as prognostication at the extremes of risk is often most relevant to decision-making yet tends to be neglected by conventional approaches (e.g., Model 1). Importantly, improvements in the tails did not come at the expense of performance in the medium-risk groups in the external cohorts, where results were broadly maintained (Figure 3).

Among sampling techniques for handling imbalance, SMOTE [14] is a well-established synthetic oversampling method for binary classification, hence a direct extension of SMOTE to our problem is not straightforward. In Appendix pp.11-15, we compare our approach to SMOTE oversampling. In our implementation of SMOTE, we balance events vs. censored within treatment groups at the global level, ignoring the prognostic structure of the cohort. From a technical perspective, SMOTE can exacerbate distortions in precisely the strata that matter most. In low-risk groups, where events are rare, it cannot generate synthetic events when few or none are observed, leaving imbalance unresolved. In high-risk groups, where events are already common, it may oversample events further (since events are globally the minority), artificially inflating their prevalence relative to censored cases. In contrast, our risk-stratified approach directly addresses imbalance across prognostic strata.

Our empirical results show that neither resampling method universally outperforms the other, but both consistently outperformed the current gold standard of training Cox models without any balancing. More importantly, both approaches (ours and SMOTE) reliably improve performance in the two tails of the external datasets – low-risk and high-risk patients. This finding, along with our sensitivity analyses (varying binning strategies and selection of prognostic factors), confirms that our observed gains in the tails are not merely statistical artifacts. Instead, the consistency across cohorts suggests that balancing, whether through our approach or SMOTE, systematically enhances model accuracy precisely where clinical decision-making is most critical. While a dedicated future study should examine in which scenarios our approach or SMOTE performs best, our current analysis already supports two conclusions: (i) balancing strategies outperform the unadjusted standard, and (ii) our approach frequently surpasses SMOTE, underscoring its practical value.

More broadly, our findings suggest that rethinking how training data are structured—not only which models are applied—may play a critical role in improving model accuracy in underrepresented subgroups across diverse healthcare applications.

# Acknowledgments

The authors would like to thank Dr. Nefeli Bampatsikou for the excellent technical support.

This research was supported by the NCI Cancer Center Support Grant P30 CA008748 (G.A.M.) and the Abdul Latif Jameel Clinic for Machine Learning in Health (D.B.).

## Tables and Figures

| Characteristic | Surgery + chemotherapy (n (%)) | Surgery alone (n (%)) | P-value (Cohen's D) |
|---|---|---|---|
| Total Patients | 1197 | 602 | |
| Age in years (IQR) | 60 (52.0-67.0) | 65.0 (58.0-73.0) | < 0.001(0.47) |
| Sex (not a prognostic factor) | | | |
| Male | 713 (59.6%) | 371 (61.6%) | |
| Female | 484 (40.4%) | 231 (38.4%) | 0.3990 |
| Median CEA in µg/L (IQR) | 8.0 (3.0-40.0) | 9.9 (3.4-44.6) | 0.3950 (0.03) |
| Median diameter of largest CRLM in cm (IQR) | 3.0 (2.0-4.0) | 3.3 (2.0-4.0) | 0.7041 (0.12) |
| Median number of CRLMs (IQR) | 2.0 (1.0-3.0) | 2.0 (1.0-3.0) | 0.2190 (0.07) |
| T category of primary tumor | | | |
| 0 | 1 (0.1%) | 4 (0.7%) | |
| 1 | 53 (4.4%) | 8 (1.3%) | |
| 2 | 203 (17.0%) | 58 (9.6%) | |
| 3 | 672 (56.1%) | 387 (64.3%) | |
| 4 | 268 (22.4%) | 145 (24.1%) | < 0.001 |
| Primary lymph node involvement | | | |
| No metastases | 495 (41.4%) | 210 (34.9%) | |
| Metastases | 694 (58.0%) | 381 (63.3%) | 0.008 |
| Primary tumor side | | | |
| Right | 563 (47.0%) | 173 (28.7%) | |
| Left | 331 (27.7%) | 238 (39.5%) | |
| Rectal | 303 (25.3%) | 191 (31.7%) | < 0.001 |
| Extrahepatic disease | | | |
| 0 | 1078 (90.1%) | 516 (85.7%) | 0.0062 |
| 1 | 119 (9.9%) | 86 (14.3%) | |
| Surgical margin status | | | |
| R0 | 1068 (89.2%) | 458 (76.1%) | |
| R1 | 129 (10.8%) | 144 (23.9%) | < 0.001 |
| KRAS mutation | | | |
| 0 | 676 (56.5%) | 398 (66.1%) | |
| 1 | 521 (43.5%) | 204 (33.9%) | < 0.001 |

Table 1: Baseline characteristics of "surgery + chemotherapy" or "surgery alone" CRLM patients before prognostic stratum matching. These are the predictor variables post-imputation which we use to train CPH models (except for "sex").

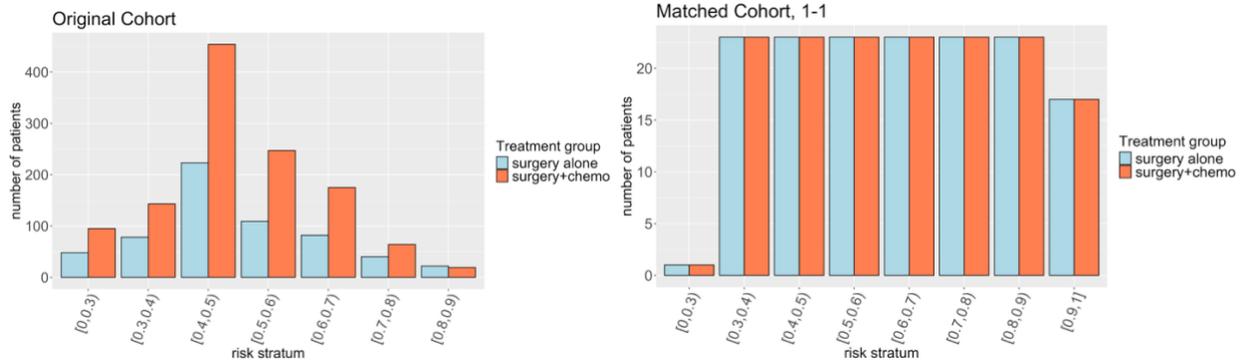

Figure 1: Prognostic risk distributions of our development cohort pre- and post-balancing. The stratification by deciles shows that the extreme tails are underrepresented in the original dataset (risk <0.3 and >0.9). *See Supplemental Figures S9, S10 for histograms related to our sensitivity analyses varying number of strata and prognostic variables*.

| Model | Harrell's C | ICI |
|---|---|---|
| 1 | 0.6144 [0.5931, 0.6356] | 0.0049 |
| **Group A ("surgery alone")** | | |
| 2A | 0.6264 [0.5918, 0.6609] | 0.0152 |
| 3A 1-1 | **0.6725 [0.6125, 0.7325]** | 0.0087 |
| 3A relaxed | 0.6619 [0.5974, 0.7263] | **0.0007** |
| **Group B ("surgery + chemotherapy")** | | |
| 2B | 0.5806 [0.5531, 0.6080] | **0.0076** |
| 3B 1-1 | 0.6519 [0.5794, 0.7245] | 0.0181 |
| 3B relaxed | **0.6993 [0.6298, 0.7688]** | 0.0305 |

Table 2: In-sample discrimination (Harrell's C) and calibration (ICI) metrics. Best scores among Group A and Group B models are highlighted in bold.

| Model | Harrell's C | 5-year AUC | ICI |
|---|---|---|---|
| 1 | 0.6115 [0.5907, 0.6336] | 0.6451 [0.6147, 0.6795] | 0.0340 |
| **Group A ("surgery alone")** | | | |

| Model | Harrell's C | 5-year AUC | ICI |
|---|---|---|---|
| 2A | 0.6201 [0.5878,0.6564] | 0.6583 [0.6079, 0.7033] | 0.0545 |
| 3A 1-1 | **0.6535 [0.6073,0.7150]** | **0.7429 [0.6719, 0.8168]** | **0.0225** |
| 3A relaxed | 0.6530 [0.5865,0.7275] | 0.7082 [0.6146, 0.7845] | 0.0305 |
| **Group B ("surgery + chemotherapy")** | | | |
| 2B | 0.5767 [0.5508,0.6043] | 0.5966 [0.5579, 0.6267] | **0.0210** |
| 3B 1-1 | 0.6392 [0.5626,0.7181] | 0.6190 [0.4693, 0.7214] | 0.0531 |
| 3B relaxed | **0.6747 [0.6057,0.7460]** | **0.6542 [5594, 0.7514]** | 0.0755 |

*Table 3: Discrimination (Harrell's C, time-dependent 5-year AUC) and calibration (ICI) metrics using bias-corrected bootstrapping. Best scores among Group A models and Group B models are highlighted in bold.*

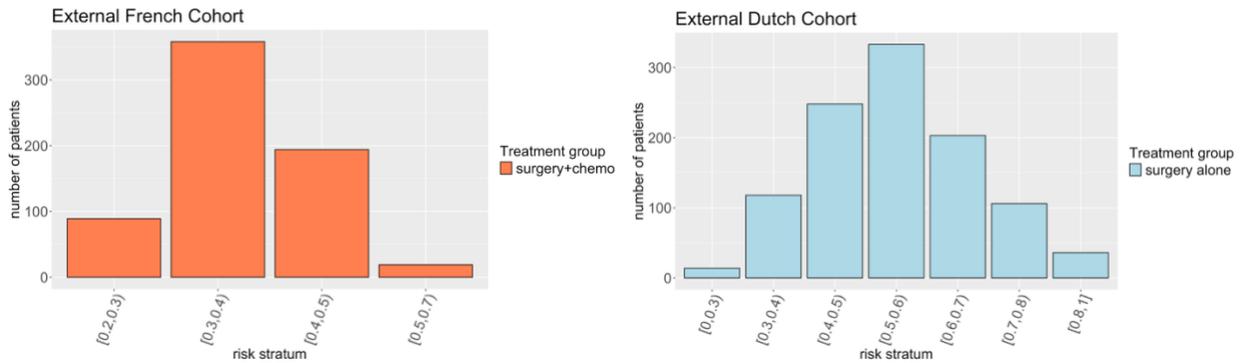

*Figure 2: Prognostic (five-year mortality) risk distribution in the French (left) and Dutch (right) datasets.*

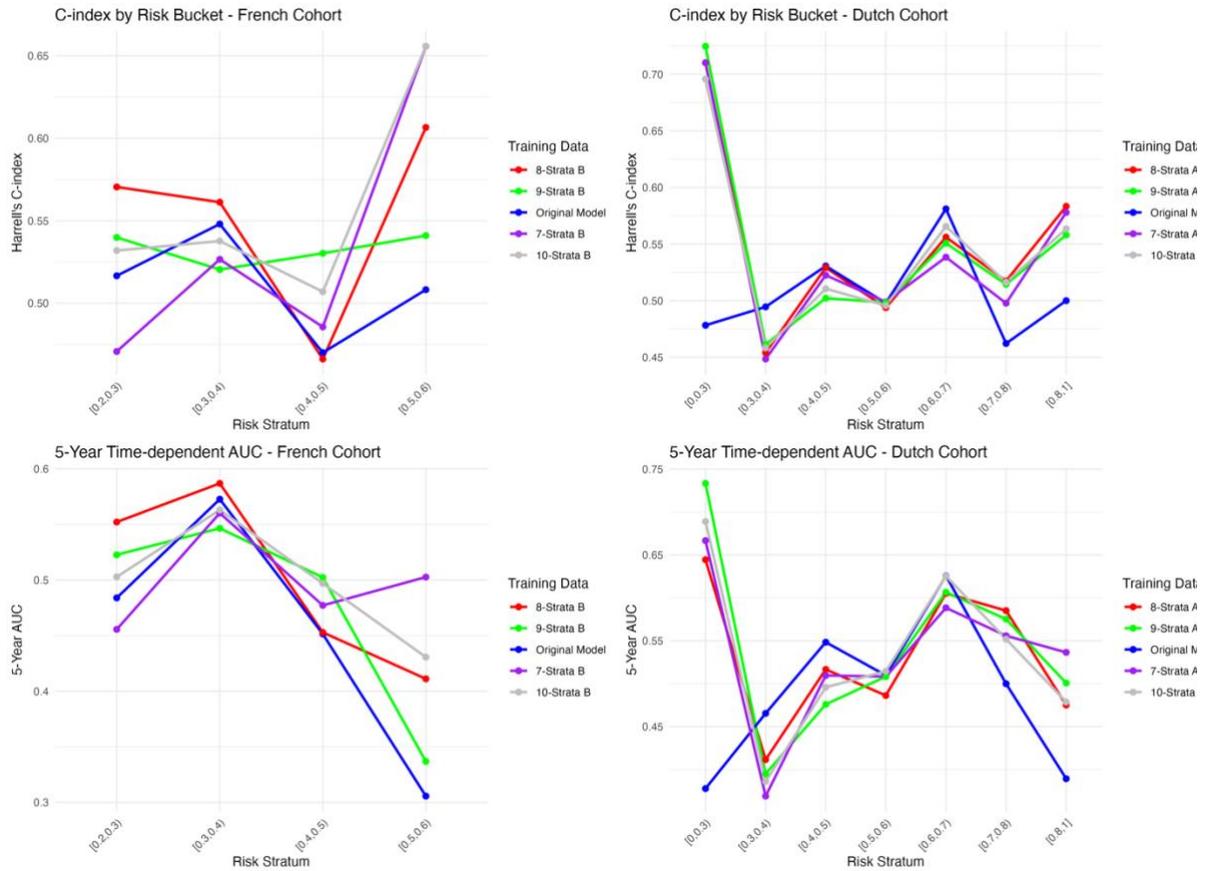

*Figure 3: Granular external validation showing per-stratum Harrell's C and time-dependent AUC indices for Models 3A and 3B, compared against Model 1, including a sensitivity analysis when varying number of strata (binning strategy) from S=7 to S=10. S=8 (red) is our chosen strategy presented in Methods.*